\documentclass[%
 aip,
 amsmath,amssymb,
 reprint,%
]{revtex4-1}

\usepackage{graphicx}% Include figure files
\usepackage{dcolumn}% Align table columns on decimal point
\usepackage{bm}% bold math
\usepackage[utf8]{inputenc}
\usepackage[T1]{fontenc}
\usepackage{mathptmx}
\usepackage{siunitx}

\usepackage{color}
\usepackage{graphicx}
\usepackage{bm}
\usepackage{amsmath}
\usepackage{amsfonts}
\usepackage{amssymb}
\usepackage{braket}
\usepackage{csquotes}
\usepackage{upgreek}
\usepackage{xcolor}
\usepackage{fancyhdr}
\usepackage{float}
\usepackage{siunitx}
\DeclareSIUnit \Vpp{\text{$\text{V}_\text{pp}$}}
\DeclareSIUnit \ohm{\text{$\Omega$}}
\usepackage{lipsum}
\usepackage{printlen}

\usepackage{import}

\makeatletter
\def\@email#1#2{%
 \endgroup
 \patchcmd{\titleblock@produce}
  {\frontmatter@RRAPformat}
  {\frontmatter@RRAPformat{\produce@RRAP{*#1\href{mailto:#2}{#2}}}\frontmatter@RRAPformat}
  {}{}
}%
\makeatother
\begin{document}

\preprint{AIP/123-QED}

\title[]{Two-phase driving of a linear radio-frequency ion trap}
% Force line breaks with \\
\author{Santhosh Surendra}
\affiliation{Physikalisches Institut, University of Bonn, Germany}
 %\altaffiliation[]{}%Lines break automatically or can be forced with \\
\email{santhosh@uni-bonn.de}

\author{Akos Hoffmann}%
\affiliation{Physikalisches Institut, University of Bonn, Germany}%\\This line break forced with \textbackslash\textbackslash

\email{ahoffmann@uni-bonn.de}

\author{Michael K\"ohl}
\affiliation{Physikalisches Institut, University of Bonn, Germany}%\\This line break forced with 
%\affiliation{%
%\\This line break forced% with \\
%}%
\email{michael.koehl@uni-bonn.de}
%\homepage{https://www.pi.uni-bonn.de/koehl/en}

\date{\today}% It is always \today, today,

\begin{abstract}
 A linear radio-frequency Paul trap is traditionally driven with one diagonal pair of electrodes grounded and the other connected to a high-voltage radio-frequency source. This method simplifies impedance matching of the voltage source to the trap. However, for several architectures it leads to increasing the axial micromotion amplitude, for example, when the capacitance between radio-frequency and end-cap electrodes is not negligible. Here, we present a technique to generate two high-voltage radio-frequency signals   \SI{180}{\degree} out of phase to drive a linear Paul trap with opposite voltages between neighbouring electrodes. Using this, we have successfully trapped and cooled a chain of Ytterbium ions in a linear radio-frequency Paul trap. 
\end{abstract}

\maketitle

\section{Introduction}

Trapped ions in radio-frequency traps are an outstanding experimental platform  to build optical atomic clocks \cite{brewer_alu_clock_2019, tofful_171yb_2024}, quantum processors \cite{moses_race-track_2023, chen_benchmarking_2024}, and quantum information nodes \cite{blinov_observation_2004,stute_tunable_2012,stute_quantum-state_2013,bock_high-fidelity_2018,kobel_deterministic_2021,  krutyanskiy_entanglement_2023}. They offer exquisite control of single-particle and entangled quantum states, and the interaction between ions and photons. Since the inception of the radio-frequency ion trap in the 1950s numerous variants of the radio-frequency ion trap, such as linear traps, ring traps, or microtraps on a chip, have been designed and built. A common challenge of all trap architectures is the delivery of electrical power to the trap. Typically, a few hundred Volt at several tens of Megahertz frequency are required for a stable operation of the ion trap.

In the radio-frequency Paul trap, ions conduct a driven oscillation -- the so-called micromotion -- at the electrical drive frequency, typically in tens of \si{\mega\hertz} range. Superimposed is a slow oscillation -- the so-called secular motion -- at a frequency typically an order of magnitude smaller. The amplitude of the micromotion is proportional to the local strength of the radio-frequency electric field and, therefore, depends strongly on the location of the ions in the trap. Even though a certain amount of micromotion is inevitable for the operation of the radio-frequency trap, an excessive amount of micromotion is detrimental for many applications. Therefore, minimization of micromotion, e.g.  by translating ions into suitable positions, is a key requirement of ion trap experiments.
	
  	\begin{figure*}[ht]
		\centering
		\includegraphics[width=0.95 \textwidth]{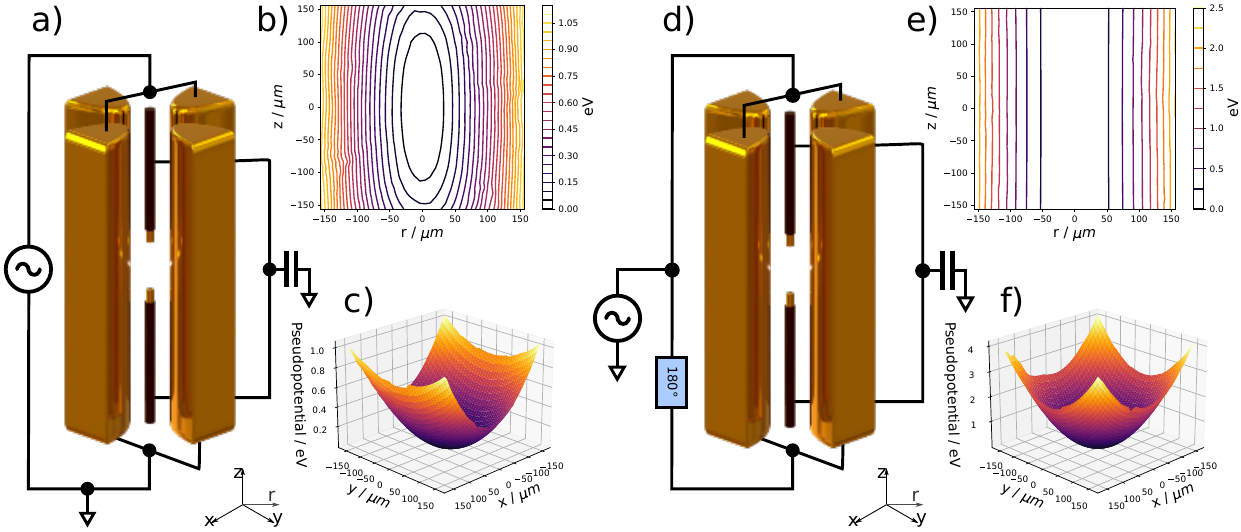}
		\caption{{Different driving schemes of a linear radio-frequency ion trap, and their pseudopotentials.} a) Single-phase driving where a diagonal pair of radio-frequency electrodes are connected to high-voltage radio-frequency source, and the other pair is grounded. b) The pseudopotential in the \textit{z-r} plane close to the trap center for single-phase drive. Observe the imballance between \textit{x-y} directions. c) The pseudopotential in the \textit{x-y} plane close to the trap center for single-phase drive. Observe the induced pseudopotential component along the trap axis (\textit{z} axis).  d) Two-phase driving scheme where the two diagonal pairs are both connected to the radio-frequency voltage source, but with \SI{180}{\degree} phase difference. In this scheme, the voltage of the two pairs oscillate about the reference ground potential. e) The pseudopotential in the \textit{z-r} plane close to the trap center for two-phase drive. f) The pseudopotential in the \textit{x-y} plane close to the trap center for two-phase drive. One can observe the deformation of the potential in the two-phase drive scheme is minimised. The pseudopotentials are numerically simulated using \cite{software_comsol} with a radio-frequency input $V_{pp} = \SI{800}{V}$ for the trap shown in Figure \ref{fig:trap_photo}.}\label{fig:drive_scheme}
	\end{figure*}

A linear radio-frequency trap incorporates four linear electrodes where typically one diagonal pair is driven by a high-voltage radio-frequency source, and the other pair is grounded, see Figure  \ref{fig:drive_scheme}a. In such a configuration -- which we refer to as single-phase drive -- the ions are trapped along a central line at which  the radio-frequency field is exactly zero. Electrically, a radio-frequency Paul trap can be considered as an open-ended conductor when driven at frequencies much lower than its self-resonance frequency (typically $>\SI{500}{\mega\hertz}$). For matching the impedance between a conventional \SI{50}{\ohm} radio-frequency voltage amplifier and the (high impedance) trap, many researches have employed helical resonators \cite{macalpine_coaxial_1959, sichak_coaxial_1955, siverns_application_2012}. A helical resonator consists of a quarter-wave coaxial resonator with its inner conductor wound into a helix to reduce its size. From the perspective of a lumped-element circuit, the helix offers an inductance in series with the capacitance of the trap, and the combined setup forms an LCR resonator \cite{wineland_experimental_1998}. When the LCR circuit is driven on resonance, one can match the impedance of the radio-frequency source to the trap. Furthermore, the helical resonator offers good filtering against phase and amplitude noise when the $Q$--factor of the resonator is high, and the voltage in the resonator becomes resonantly enhanced.

However, only infinitely extended and perfectly parallel electrodes feature a line of exactly zero radio-frequency electric field. In realistic traps, additional end-cap electrodes confine the ions along the symmetry axis, and these electrodes are supplied with DC voltages and capacatively grounded for the radio-frequency electric field (see Figure \ref{fig:drive_scheme}a). The induced radio-frequency voltage on the DC-electrodes alters the radio-frequency potential in the trap, and in particular, it adds a small axial electric field component along the symmetry axis of the trap. This electric field component induces micromotion along this axis, which we refer to in this article as axial micromotion.

The axial micromotion in a single-phase-driven ion trap increases when the end-cap separation decreases owing to stronger potential alteration by the electrodes. In the limit where the distance $z_0$ between ion and end-cap electrode is similar to or smaller than the distance $\rho_0$ between ion and radio-frequency electrode, the axial micromotion amplitude becomes comparable to the radial micromotion amplitude. Hence, the ion trap can only accommodate a single cold ion in a nearly micromotion-free location. The pseudopotentials of such a trap along the radial and axial planes are shown in Figure \ref{fig:drive_scheme} b) and c). One can observe the distortion of the potential in the radial plane, and the pseudopotential due to the induced electric field along the axis of the trap. %We would like to work in such a limit by using metallized fiber optical cavity as end caps \cite{pfeifer_achievements_2022}.

A promising way to reduce the axial micromotion is by using a two-phase electrical drive of the trap electrodes (see Figure \ref{fig:drive_scheme}d). Here, the grounding of a diagonal pair of electrodes is replaced by a second radio-frequency drive with a phase difference of \SI{180}{\degree}. With such a drive, one  observes that the pseudopotential in the radial plane is not distorted, and the pseudopotential due to induced electric field along the trap axis is minimised [see Figure \ref{fig:drive_scheme} e) and f)]. Two-phase driving of a linear radio-frequency trap has been implemented in some experiments before by using a center-tapped helical resonator \cite{rosnagel_single-atom_2016, wubbena_controlling_2014}, meander trace on a circuit board \cite{chen_ticking_2017}, direct driving via a balun transformer \cite{thomsen_design_2020}, and by direct driving using two phase-adjustable signal generators \cite{zipkes_trapped_2010thesis}. Using superconducting electronics, low-noise ion traps with very high quality factors  have been built and operated \cite{stark_ultralow-noise_2021}. The challenging aspect of the two-phase drive is that any  phase difference other than 0 or ($180^\circ$) between opposite (neighbouring) electrodes leads to  excess micromotion \cite{berkeland_minimization_1998}.

%{\color{blue} Are you sure we dont have to include a paragraph to describe the cons of alternate techniques? and, are we sure to not explicitly mention something like "there are not many publications for this, and hence we are filling some gap in literature..... considering this is an instrumentation journal" }

%{\bf Moreover, center tapping a helical resonator is challenging from a fabrication standpoint, where the input power can drop across the resonator and not the trap electrodes if machined imperfectly. Meander trace on PCB boards are limited in the amount of voltage they can handle, and their physical dimension become very large for frequencies of around \SI{20}{\mega\hertz}. They also offer higher losses, and hence lower Q-factor. This reduces the ability of the resonator to reject noise. Direct driving using DDS and Balun transformers are also limited in the voltage amplitude, and their noise rejection. -- MK: I am not 100\% sure we need this paragraph. If you prefer, important elements of this could be shortened/condensed into the previous paragarph.}

In this work, we have designed, simulated, and implemented a two-phase resonator  providing two radio-frequency outputs with opposite phase. With this, we have achieved trapping of linear chains of Ytterbium ions at secular radial trapping frequencies above \SI{1.2}{\mega\hertz}.
	
\section{Resonator theory}
	
\begin{figure*}[ht]
		\centering
		\includegraphics[width=0.85 \textwidth]{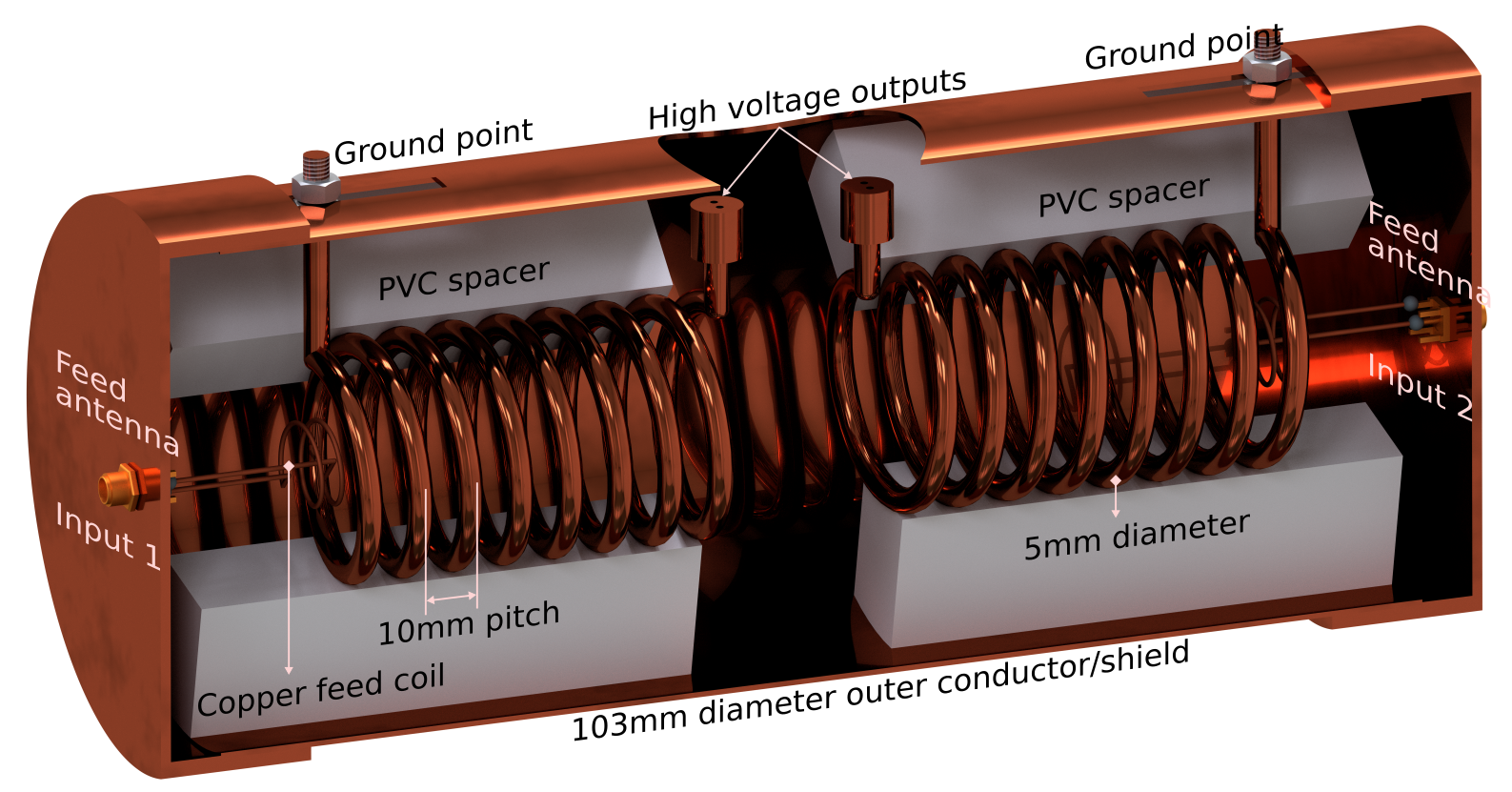}
		\caption{{Design of the two-phase helical resonator.} The resonator consists of two helical inner conductors in a copper shield, but with opposite helicity. The two opposite ends are grounded to the shield using thread and nuts, and the free ends offer the two outputs for the trap. Radio-frequency power can be fed into the system on either side using the feed antennas. The large coils are held in place using PVC spacers.}\label{fig:resonator_design}
	\end{figure*}

	Our overall strategy to obtain two voltage outputs with opposite phases is to construct a helical resonator with a second helix with opposite helicity and couple both to form a common  inductance in series with the trap, as shown in Figure \ref{fig:resonator_design}. We ground the two helices at the far ends to the shield, such that each of them can accommodate standing waves of the electromagnetic field. The currents in the helices couple them inductively with each other. Furthermore, the small physical separation of the two helices creates a noticeable capacitance between them, which enhances the mutual coupling. The two open ends are the two high-voltage outputs to drive the ion trap.

\begin{figure}[h]  
	\centering
	\includegraphics[width=0.5 \textwidth]{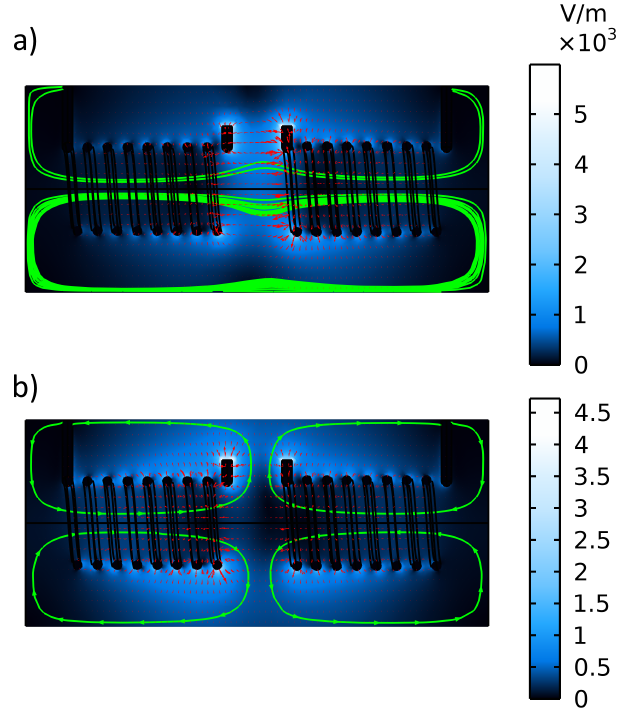}
	\caption{{Lowest two frequency eigenmodes of the electromagnetic field inside the coupled resonator.} The eigenmodes were simulated using a FEM software \cite{software_comsol}, and one cross-section of the electromagnetic field is shown here. a) The lower-frequency mode is the asymemetric mode with the two outputs of the resonator oscillating out of phase. It is the desired mode to operate the radio-frequency ion trap. b) The higher-frequency mode is the symmetric mode where the two outputs of the resonator oscillate in phase. The heat map show the absolute value of the electric field, for an arbitrary starting condition of the solver. Red arrows show the electric field direction, and the green streamline shows the magnetic field distribution for the two modes.}
	\label{fig:comsol}
\end{figure}

	Our resonator system accommodates several eigenmodes of the electromagnetic field with different electric and magnetic field distributions. In order to analyse the modes, we have simulated our resonator using a FEM solver \cite{software_comsol}. The electric and magnetic field distribution of the two lowest eigenmodes are shown in Figure \ref{fig:comsol}. The mode with lower frequency has an output with opposite voltages at endpoints of the two helices and, therefore, is the desired mode to drive the trap.

\begin{figure}[h]
	\includegraphics[width=0.5 \textwidth]{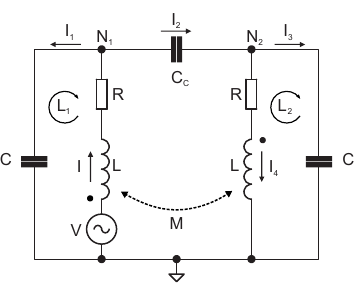}
	\caption{{A simple lumped-element model of the coupled resonator system.} The two inductors of inductance L are coupled at the opposite ends due to the opposite helicity by an inductive coupling M. Furthermore, there is a coupling capacitance $C_c$ that connects the two outputs ($N_1$, $N_2$) of the two resonators.} 
	\label{fig:lumped_circuit_simple}
\end{figure}

	With the knowledge of the electric field distribution inside the double-helical resonator, we now turn our attention to modelling the electrical behaviour of the whole system. We begin with a lumped-circuit model using coupled LCR tank circuits as shown in Figure \ref{fig:lumped_circuit_simple}. The double-helical resonator is replaced by the two equal inductances $L$. The parasitic self-capacitances of the coils are represented by $C$ and are assumed to be equal. This also includes the parasitic capacitance of the helices with the outer conductor/shield. Coupling between the helices is facilitated by the inductive coupling $M$ and the capacitive coupling ${C_c}$. The latter includes the capacitance between the two coils, the capacitance due to the wires between the resonator and the vacuum chamber, the vacuum feed through, and the ion trap. Since all of these are in parallel to each other, we can sum them all up into one coupling capacitance. The ohmic resistances $R$ describe the loss due to resistance of all copper conductors, approximately \SI{0.1}{\ohm} at tens of \si{\mega\hertz} for skin depth of approximately \SI{12}{\micro\meter}.	
	
We analyse the double-helical resonator circuit using Kirchhoff's laws. With a voltage drive $V(t) = V_0\sin \omega t$ we obtain for the two loops
	\begin{align}
		\ddot{I} + \frac{M}{L}\ddot{I}_4 + \frac{R}{L} \dot{I} + \frac{1}{LC} I_1 = \frac{V_0 \omega}{L} \cos{\omega t} \label{KVL:loop1}
	\end{align}
	\begin{align}
		\ddot{I}_4 + \frac{M}{L}\ddot{I} + \frac{R}{L} \dot{I}_4 - \frac{1}{LC} I_3 = 0. \label{KVL:loop3} 
	\end{align}
We rewrite the two differential equations in terms of the symmetric and the asymmetric currents $I_s = I_1 + I_3$ and $I_a = I_2$, respectively, and obtain two independent differential equations
		\begin{align}
			\ddot{I}_a + \Gamma_a \dot{I}_a + \omega_a^2 I_a &= \frac{V_0 \omega}{\left(L + M \right) \left(2 + \frac{C}{C_c}\right)} \cos{\omega t}\label{eq:asy}\\
			\label{eq:symmetric}
			\ddot{I}_s + \Gamma_s \dot{I}_s + \omega_s^2 I_s &= \frac{V_0 \omega}{\left(L-M\right)} \cos{\omega t}
		\end{align} 		
These equations describe two harmonic oscillation modes of the coupled LCR tank circuits. Here, $\Gamma_i$ are the line widths of the resonances, and $\omega_i$ are the resonance frequencies. The quality factor is $Q_i = \omega_i/\Gamma_i$ for the two resonances, and their respective expressions are tabulated in Table \ref{tab:resonances}. The asymmetric resonance, where the two LCR tank circuits oscillate out of phase, has the lower resonance frequency, and the symmetric resonance, in which the two tank circuits oscillate in phase, has the higher resonance frequency. %Thus, we use the lower energy antisymmetric mode to drive the trap with two radio-frequency outputs with phase difference of \SI{180}{\degree}.
	%\end{widetext}

	\begin{table}[h]
	\centering
	\caption{{The normal modes of the coupled LCR resonator system.}}
	\label{tab:resonances}
	\begin{tabular}{ |c|c|c| } 
		\hline
		& Symmetric resonance & Asymmetric resonance \\ 
		\hline
		$\omega$ & $\sqrt{\frac{1}{\left(L-M\right)C}}$ & $\sqrt{\frac{1}{\left(L+M\right) \left(2{C_c} + C\right)}}$ \\ 
		Q & $\frac{1}{R} \sqrt{\frac{\left(L-M\right)}{C} }$ & $\frac{1}{R} \sqrt{\frac{\left(L+M\right)}{\left(2{C_c} + C\right)}}$ \\ 
		\hline
	\end{tabular}
\end{table}

\begin{figure*}[ht]
	\centering
	\includegraphics[width=0.95\textwidth]{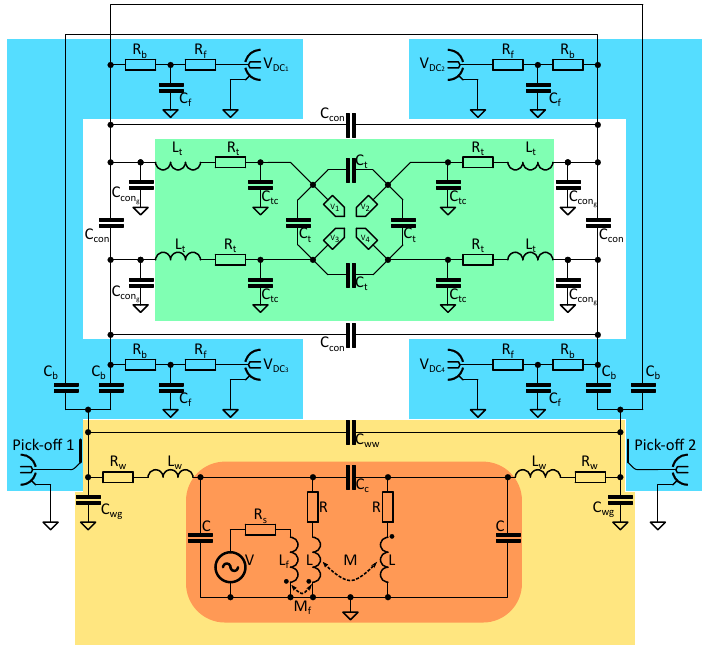}
	\caption{{The complete equivalent circuit of our system.} The orange region contains the two-phase helical resonator, similar to Figure \ref{fig:lumped_circuit_simple}. Here we include one of the feed antennas with an antenna-resonator coupling given by $M_f$. The yellow region includes the resonator with the silver-coated wires providing the two-phase outputs to the bias tee. The blue region depicts the bias tee with the four independent DC inputs, and the two pick-off antennas. The pick-off antennas are two open ended PCB traces with a SMA connector (see Figure \ref{fig:bias_tee_design}). This region contains all the real electronic components that are soldered onto the PCB. All other inductors, capacitors, and resistors are parasitic components describing various connecting wires and the ambience. The green region contains the ion trap with the vacuum chamber and its feed-through. The white region between the bias tee and the vacuum chamber shows the connection between the bias tee and the vacuum feed-through. \label{fig:lumped_circuit_full}}
\end{figure*}

The entire system, however, is a more complicated assembly that includes further parasitic capacitances and inductances. In order to incorporate these, as the next step, we solve the system numerically using \cite{software_ltspice}. We use a simplified Heaviside transmission line model with a series LCR circuit describing every conductive component. This method has been shown to predict the behaviour of such systems better by including the parasitics of the connecting wires and traces \cite{thomsen_design_2020}. Our best version of the equivalent circuit diagram is shown in Figure \ref{fig:lumped_circuit_full}.

At the center of Figure \ref{fig:lumped_circuit_full} is the linear ion trap with the electrodes and their voltages represented by ${V_i}$. Electrically, we consider each of the trap electrodes to be inclusive of their connecting wires inside the vacuum chamber and the vacuum feed-through. The capacitance between the neighbouring pairs of electrodes is represented by $C_t$, and is measured to be \SI{1.2 \pm 0.5}{\pico\farad} up to \SI{10}{\kilo\hertz} using a LCR meter from outside the vacuum chamber. The vacuum chamber, the optical table, and all metallic mounting structures are grounded and act as the ground potential reference in the circuit model.  $C_{tc}$ represents the capacitance between the electrodes on the one hand and the mounting structure and the vacuum chamber on the other hand. We estimate it to be \SI{1.9}{\pico\farad} for a wire length of  \SI{10}{\centi\meter}. $L_t$ represents the self inductance of the electrodes and is estimated below. $R_t$ is the resistance of electrodes and connecting wires at drive frequency, and is calculated to be \SI{0.05}{\ohm} at \SI{30}{\mega\hertz}. 

Self inductance of a straight wire of radius $r$, and length $\Lambda$ is given by \cite{rosa_self_1908}:
\begin{align}%
	L_w = \frac{\mu_0}{2\pi} \Lambda \left[ \ln\left(\frac{2\Lambda}{r}\right) -1 \right]%\si{\henry}
	\label{eqn:Lw}
\end{align}
giving an estimate for $L_t$ to be \SI{100}{\nano\henry}.

The blue blocks in Figure \ref{fig:lumped_circuit_full} are the bias tee used to add DC voltages to the radio-frequency electrodes. $C_b$ is the series capacitor of the bias tee, and $R_b$ is the parallel resistor connecting the high-voltage part of the circuit with the DC voltage sources. They also offer a parasitic capacitance with ground shown by $C_{con_g}$, which is harder to estimate due to more complex geometry. We use a value of \SI{1}{\pico\farad} in the model, because an LCR meter measures approximately that much for an identical copper wire above a flat grounded surface.

The two wires that transfer high voltage from the helical resonator to the bias tee are shown by an equivalent LCR circuit consisting of an inductance $L_w$, resistance $R_w$, a capacitance between them given by $C_{ww}$, and a capacitance with the ground given by $C_{wg}$. The two wires are separated by around \SI{20}{\milli\meter}, and are about \SI{8}{\centi\meter} above the optical table. The capacitance between the two parallel wires is given by  \cite{jackson_classical_1999}
\begin{align}
	%C_{ww} = \epsilon_0 \frac{ \pi \Lambda}{\cosh^{-1}\left(\frac{d}{2r}\right)}\label{eqn:Cw},
	C_{ww} = \epsilon_0 \frac{ 2\pi \Lambda}{\cosh^{-1}\left(\frac{d^2}{2r^2}-1\right)}\label{eqn:Cw},
\end{align}
where $\Lambda$ is the length of the wires, $r$ is the radius of the wires, and $d$ is the distance between them. This is estimated to be \SI{1.1}{\pico\farad} in our case. The capacitance between each wire and the optical table is estimated using  \cite{clayton_analysis_2008}
\begin{align}
	C_{wg} = \epsilon_0 \frac{ 2\pi \Lambda}{\cosh^{-1}\left(\frac{h}{r}\right)},
\end{align}
where $h$ is the height of the wire from the table surface. This is around \SI{1.4}{\pico\farad}. The resistance due to the wires is $R_w$ and is estimated to be \SI{100}{\milli\ohm}. The self inductance of the wires is $L_w$ and is calculated to be \SI{200}{\nano\henry}. We do not include a mutual inductance between the two parallel wires since it is negligible compared to the mutual inductance between the two helical coils of the resonator described bellow. 
 
The circuit equivalent of the helical resonator is shown in the orange block in Figure \ref{fig:lumped_circuit_full}. For a coil diameter $D_c$, height $b$, pitch $\tau$, and outer conductor/shield diameter $D_s$, the self capacitance of a coil is given by \cite{siverns_application_2012}
 \begin{align}
 	C_c \approx  11.26 \tfrac{\text{pF}}{\text{m}}\; b +  8\tfrac{\text{pF}}{\text{m}}\;D_c + 27 \tfrac{\text{pF}}{\text{m}}\sqrt{\tfrac{D_c^3}{b}} \simeq 2.1\,\text{pF}.
 \end{align}
The capacitance between a coil and the shield is approximately \cite{macalpine_coaxial_1959}
\begin{align}
 	C_s \approx 29.53 \tfrac{\text{pF}}{\text{m}}   \frac{b}{\log\frac{D_s}{D_c}} \simeq 2.6 \,\text{pF}
 \end{align}
 The capacitances $C_c$ and $C_s$ are in parallel to each other and form the total capacitance of each resonator arm $C = C_c + C_s$. The inductance of a short coil inside a shield is \cite{macalpine_coaxial_1959}
\begin{align}
 	L_c \approx 0.984\,\tfrac{\mu\text{H}}{\text{m}}\,  \frac{D_c^2 b}{\tau^2} \left[1-\left(\tfrac{D_c}{D_s}\right)^2 \right] \simeq 0.9\,\mu\text{H},
 \end{align}
where we have used the inner diameter of the coil. The mutual induction $M$ between two coils is defined as the magnetic flux through coil 2 produced by coil 1 with a unit current. One can observe from the numerical simulation of the electromagnetic field distribution inside the resonator (see Figure \ref{fig:comsol}) that the magnetic field lines that enter a face of the helices do not escape them laterally, but exit through the opposite face. Hence, we model the helices as a single-turn current loop of the same diameter and separation as the helices, and numerically calculate the magnetic flux through one loop due to a current of \SI{1}{\ampere} in the other one. For the loop diameter of $D_c=\,$\SI{42}{\milli\meter}, and an axial separation of $x=\,$\SI{3}{\centi\meter} (see Table \ref{tab:helical_resonator}), we estimate the inductive coupling coefficient $\kappa \approx 0.03$. Here $\kappa = M/\sqrt{L_1L_2}$ where $M$ is the mutual induction, and $L_i$ are the two inductances. Furthermore, we model the coupling capacitance $C_c$ as a capacitance formed by two planar rings of thickness $\phi$ and diameter $D_c$, separated by a distance $x=$\,\SI{3}{\centi\meter} (see Table \ref{tab:helical_resonator}). This gives us an estimate of $C_c \approx \SI{0.2}{\pico\farad}$. R is the resistance of the helical coils at \SI{30}{\mega\hertz}, and is estimated to be \SI{0.1}{\ohm}.
 
The radio-frequency source in the simulator has a $R_s = \SI{50}{\ohm}$ internal resistance, and the feed coil has an inductance of $L_f = \SI{500}{\nano\henry}$ representing the real device. With these values, we optimize the coupling $M_f$ between the feed coil and the resonator coil in order to minimize the scattering parameter S11 at the lower-frequency resonance. In reality, this process is performed in the lab by modifying the shape and pitch of the feed coil, and by translating it inside the resonator. This allows one to change the mutual inductance $M_f$ and hence perform impedance matching, see \cite{siverns_application_2012} for the details of this process.

\begin{figure}[h]
\includegraphics[width=0.5 \textwidth]{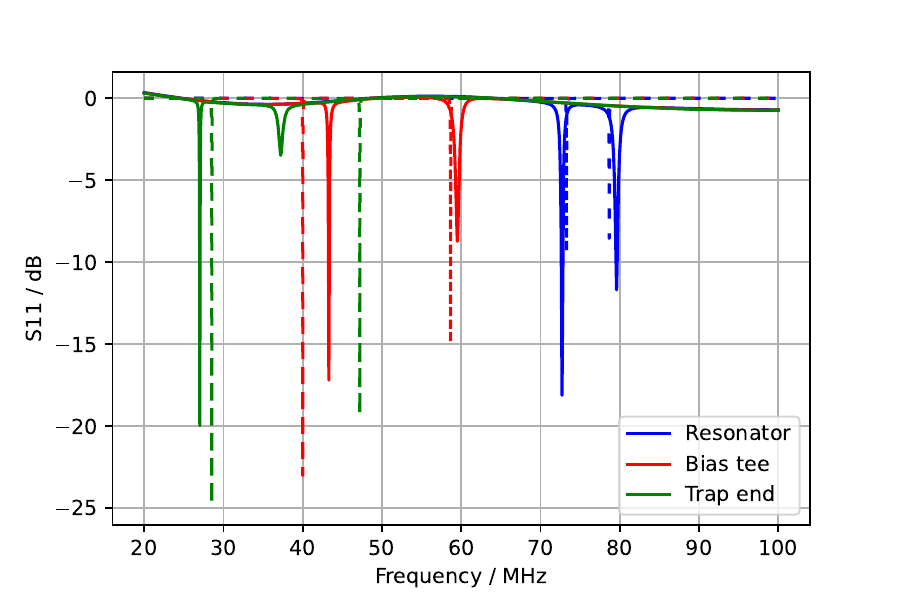}
\caption{Measured vs. calculated frequency response. Solid lines are the measured scattering parameter S11 using a vector network analyser (Pico VNA 108), and the dashed lines are the numerically simulated results from the circuit model in figure \ref{fig:lumped_circuit_full}. The blue lines show the response of the bare resonator. Red lines show the response when the bias tee is connected. Green data shows the total response of the system with the trap.} \label{fig:resonance_shift}
\end{figure}

Figure \ref{fig:resonance_shift} shows the comparison between the measurement and the simulation results of the above model. One observes that the resonances shift when adding the bias tee, and the trap follows to within \SI{10}{\percent} to the circuit model for the lower frequency resonance. We have also performed a Monte-Carlo analysis of the sensitivity of the circuit to independent variations of the numerical values of the components. To this end, we have introduced random variations of up to  \SI{10}{\percent} on all electronic and parasitic components and computed the resonance frequency. This modelling predicts a standard deviation of \SI{0.58}{\mega\hertz} for the lower resonance frequency, and a phase difference between %two adjacent trap electrodes of $\Delta \phi_\text{adj} = 179.86 \pm 0.01 \si{\degree}$, and 
opposite trap electrodes of $\Delta \phi_\text{opp} = 0 \pm 0.00035 \si{\degree}$ on resonance.  %It is important to note that the excess-micromotion is sensitive not to $180\si{\degree} - \Delta \phi_\text{adj}$, but to the phase difference between the two adjacent electrodes -- i.e. $\Delta \phi_\text{opp}$ 
%. This is something that is very challenging to measure experimentally, and is minimised by making sure all connecting wires from the bias-tee to the trap end are symmetric and are of the same length, as described in \cite{siverns_application_2012}.

\section{Resonator design}

The resonator enclosure is built from copper in order to reduce resistive losses, and PVC spacers to hold the coils in space. PVC was used because it is easy to machine, however, one could potentially achieve lower losses with \text{PTFE} \cite{siverns_application_2012}. The helices are made from copper wire with a diameter of $\phi = \SI{5}{\milli\meter}$ and they are produced by winding the wire onto a cast. The coils have a winding pitch of $\tau = \SI{10}{\milli\meter}$, and a diameter of $D_c = \SI{42}{\milli\meter}$. We directly machined a thread onto the grounding end of the coil, and use stainless steel nuts to electrically contact them to the two slots in the outer conductor of the resonator (see Figure \ref{fig:resonator_design}). Two silver-plated copper wires transfer the voltage from the two outputs of the resonator to the high-voltage bias tee, and then the ion trap. Power can be fed into the resonator from either side using feed coils. Since the system of the two feed coils and the two main coils is coupled,  we observe a strong deterioration of the resonance if we connect both of feeds simultaneously to \SI{50}{\ohm} terminated radio-frequency amplifiers. Hence, we drive the system using a single feeding coil while keeping the other one open ended. Fine-tuning of the resonance frequency is achieved by moving the two helical conductors inside the shield with respect to each other. This changes the coupling and hence the resonance frequency of the two modes. Dimensions of the parts are given in Table \ref{tab:helical_resonator}.
\begin{table}[h]
    \centering
    \caption{Design values of our resonator.}
    \label{tab:helical_resonator}
    \begin{tabular}{|l|c|} 
        \hline
        Parameter & Value \\ 
        \hline
        Coil diameter $D_c$ & 42\,\text{mm}\\
        Coil thickness $\phi$ & \SI{5}{\milli\meter} \\ 
        Coil pitch $\tau$ & \SI{10}{\milli\meter} \\ 
        Winding pitch angle & $\approx \SI{10}{\degree}$ \\ 
        Number of turns $n$ & 8 \\ 
        Coil-coil separation $x$ & \SI{3}{\centi\meter} \\
        {Shield inner diameter $D$} & \SI{103}{\milli\meter} \\ 
        Shield length & \SI{20}{\centi\meter} \\
        {Feed diameter} & $\approx \SI{15}{\milli\meter}$ \\
        Feed pitch & $\approx \SI{4}{\milli\meter}$ \\
        Feed thickness & \SI{1}{\milli\meter} \\
        {Feed antenna turns} & 2 \\
        \hline
    \end{tabular} 
\end{table}

\begin{figure}[h]  
\centering
    \includegraphics[width= 0.5\textwidth]{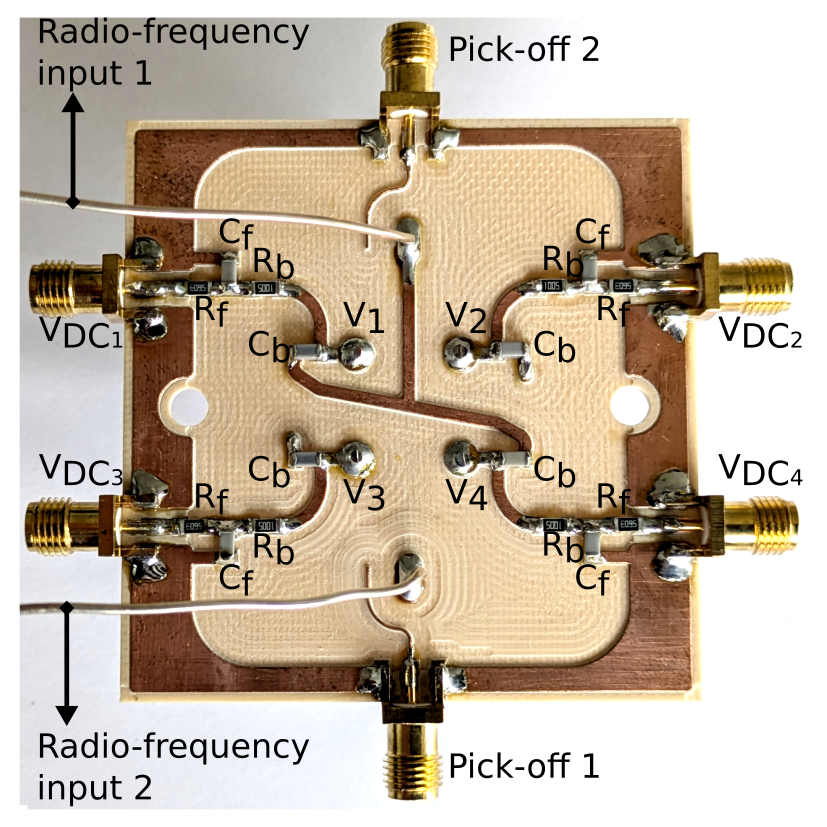}
\caption{{The high voltage bias tee}. It is fabricated on a Rogers 4003C circuit board with two silver-coated copper wires connected to the two outputs of the two-phase helical resonator. These bring in the radio-frequency high voltage from the resonator to the bias tee. The PCB trace carrying the two phases split into two paths symmetrically on the two sides of the PCB (only one can be seen here), and are connected to the four DC inputs via a bias capacitor $C_b$ and bias resistor $R_b$. The mixed DC and radio-frequency voltages $V_i$ are connected to a vacuum feed-through. The two voltage pick-off antennas can be seen at the top and bottom of the figure for monitoring the circulating voltage in the resonator-trap system.}	
\label{fig:bias_tee_design}
\end{figure}

In order to bias the four radio-frequency electrodes with independent DC voltages for excess micromotion compensation, we have fabricated a bias tee that can be plugged in between the helical resonator and the vacuum feed-through. The bias tee is fabricated on a Rogers 4003C board of thickness \SI{1.52}{\milli\meter} (see Figure \ref{fig:bias_tee_design}). All high voltage traces are of width \SI{1.42}{\milli\meter}. This value was a compromise between the trace resistance and the proximity to the other electrodes. We have kept the high-voltage ends at least \SI{3}{\milli\meter} from each other to avoid electric discharge. We observed sparking on the surface of the PCB at the shortest distance during the first trial and never again. We attribute this to thin leftover copper or residue from fabrication process. 

During the design of the PCB, the radio-frequency trace lengths were minimized, and all convex corners were smoothened to reduce resistive and radiative losses. The board also includes two pick-off antennas close to the two high-voltage traces. These allow us to monitor the circulating voltage on an oscilloscope by coupling out $\approx 0.24 \%$  of the radio-frequency voltage on the high-voltage ends. The bias circuit is built with a series high-voltage capacitor isolating the DC from the resonator, and a \SI{10}{\mega\ohm} resistor isolating the DC voltage sources from the induced high-voltage radio frequency. When capacitors are added in series with the to outputs of the resonator as shown in Figure \ref{fig:bias_tee_design}, the voltage drop across them is much smaller than the voltage drop across the two outputs of the resonator. This can be seen by using Kirchhoff's voltage law. Thus, by using a serial capacitance much larger than $C_c$, we can retain maximum radio-frequency voltage drop across the trap, while minimising losses. We further add a low pass filter using the resistor $R_f$, and capacitor and $C_f$ to further protect the DC voltage source from induced radio frequency. All bulk components that are used are tabulated in table \ref{tab:electronic_components}.

\begin{table}[h]
    \centering
    \caption{{Components used in the high-voltage bias tee.}}
    \label{tab:electronic_components}
    \begin{tabular}{ |l|c|c| } 
        \hline
        Component &  Value & Type \\ 
        \hline
        PCB & \SI{1.52}{\milli\meter} thick & Rogers 4003C\\ 
        RF trace width &  \SI{1.42}{\milli\meter} & Copper\\ 
        Series bias capacitor $C_b$ &  \SI{3.3}{\nano\farad} & 1206 NP0 \SI{630}{\volt} \\ 
        DC bias resistor $R_b$ &  \SI{10}{\mega\ohm} & 1206 \SI{1}{\percent} thin film\\ 
        DC bias filter resistor $R_f$ &  \SI{560}{\kilo\ohm} & 1206 \SI{1}{\percent} thin film\\ 
        DC bias capacitor $C_f$ &  \SI{220}{\pico\farad} & 1206 NP0 \SI{50}{\volt}\\ 			
        \hline
    \end{tabular} 
\end{table}

\section{The ion trap}

\begin{figure}[h]
    \includegraphics[width=0.45 \textwidth]{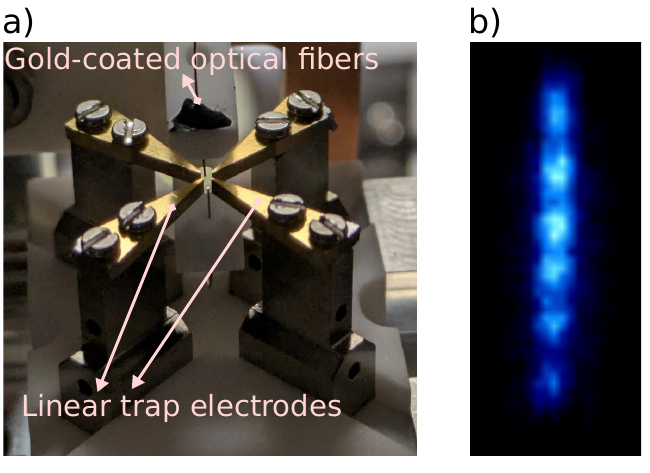}
    \caption{a) Linear Paul trap with gold-coated optical fibers as end-cap electrodes. The four radio-frequency electrodes that are closest to the ions are gold coated and the assembly is held  by a ceramic holder. b) {Chain of laser cooled and trapped Ytterbium ions.}}
    \label{fig:trap_photo}
\end{figure}

% - Ion - photos
Our radio-frequency ion trap is shown in Figure \ref{fig:trap_photo}. The four radio-frequency electrodes are fabricated from stainless steel, and the parts closest to the ions are further coated with gold to reduce surface oxide layers near the ions. All four radio-frequency electrodes are connected to independent high-voltage vacuum feedthroughs. The end-cap electrodes are \SI{125}{\micro\meter} diameter optical fibers with \SI{100}{\nano\meter} gold coating on the whole end facet. We have used \SI{5}{\nano\meter} of chromium as an adhesion layer between glass and gold. The separation of the ion from the radio-frequency electrodes and endcap electrodes are both {400 $\pm$ 40}\si{\micro\meter}. The protective layer of the fiber is made of copper, hence acting as a conductor for DC voltage. %We note here that the current implementation does not include a fiber cavity, and the fibers' end facets are completely covered in gold. 	

For a perfectly machined and aligned ion trap, one expects no induced radio-frequency voltage on the end caps due to two-phase drive. However, we observe around \SI{0.016}{\percent} of the voltage from the radio-frequency electrodes coupled onto them. This is measured at the vacuum feedthrough. We attribute this to a potential misalignment and a small amplitude imbalance between the two phases. We have added a \SI{10}{\hertz} low-pass filter at the endcap electrodes' vacuum feedthrough to protect the dc voltage sources from this small radio frequency pickup.

%\section{Results}

\begin{figure}[h]
    \centering
    \includegraphics[width=0.48 \textwidth]{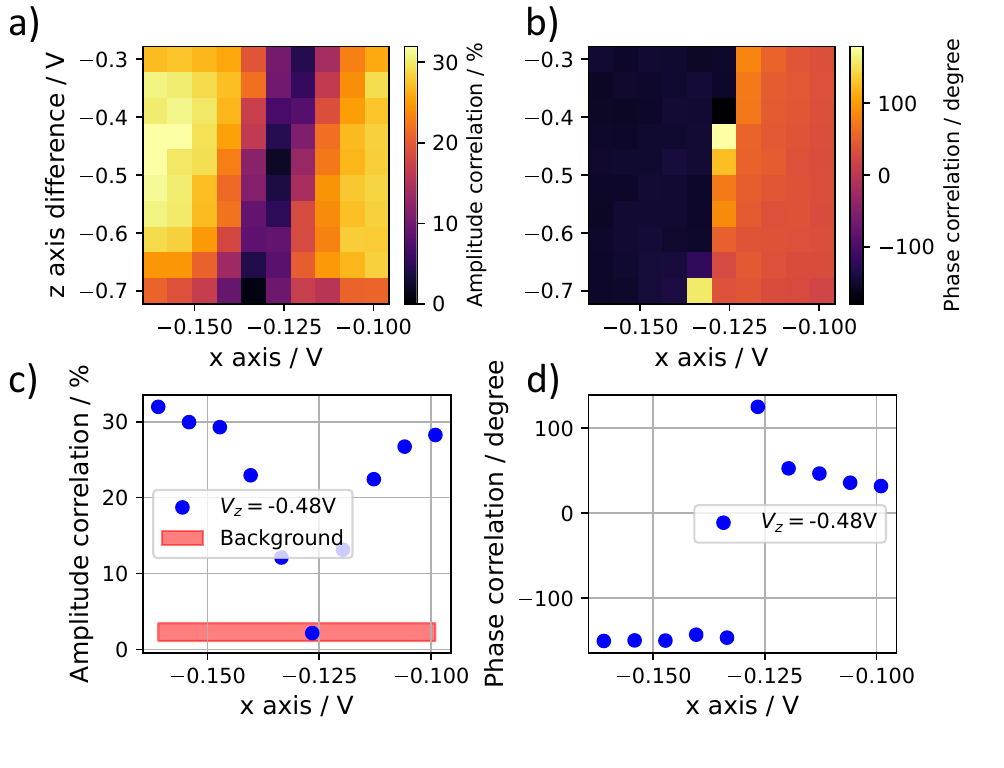}
    \caption {a) Amplitude correlation signal and b) phase correlation signal of the photon Doppler correlation measurement along x-z plane for a laser beam in y-z plane with \SI{35}{\degree} to the y axis. Each data point is a measurement performed over \SI{0.5}{\second}. The amplitude of Doppler correlation remains at its minimum as the ion is scanned across the axis of the trap by changing the end-cap voltage difference (denoted by z axis difference). The slight shift of the correlation minima along the axis is due to the misalignment of the end-cap electrodes during the vacuum bake out that cause a small deformation of the coordinate axes. c) and d) represent the central horizontal cut from a) and b), respectively. The background correlation is an average $2\sigma$ width of 50 measurements with no ions trapped.   }
    \label{fig:axial_micromotion}
\end{figure}

We have successfully trapped up to seven Ytterbium ions in a linear chain. The ion number is limited by the size of the cooling laser beam. We have observed trapping lifetime in excess of two hours while laser cooling with a single re-pumping laser at 935\,nm. Excess-micromotion  detection is performed using the photon-Doppler correlation measurement \cite{berkeland_minimization_1998}. %To this end, we obtain the timestamps of the ion fluorescence on a time tagger and calculate the amplitude and phase correlations of the photon detection times with the drive frequency. 
As an example, Figure \ref{fig:axial_micromotion} shows a typical Doppler correlation measurement for 2 ions when moved in the $xz$ plane for a laser beam in the $yz$ plane with an angle of \SI{35}{\degree} with the $y-$axis. The correlation remains minimised at its lowest value even when applying a voltage difference of  \SI{0.4}{\volt} difference across the endcap electrodes (biased at  \SI{1}{\volt}). This corresponds to a micromotion-free displacement of the ion over \SI{65}{\micro\meter}. The background correlation corresponds to \num{2} $\pm$ \num{1}\si{\percent} for the amplitude and no significant correlation for the phase. These are due to the background stray light in the lab when no ions are trapped.%\num{-5.72} $\pm$ \num{117.27} \si{\degree} for the phase {\bf this error bar is pointless. What does it even mean?}-- which is due to the background stray light in the lab when no ions are trapped.}

\begin{figure}[h]
    \centering
    \includegraphics[width=0.45 \textwidth]{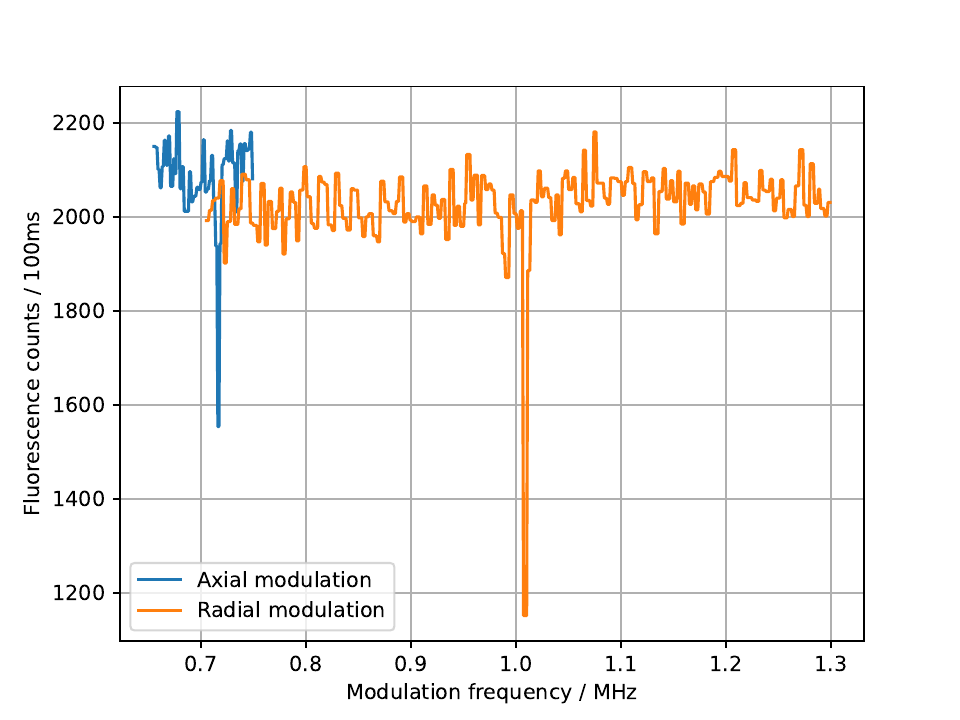}
    \caption {Trap modulation scan for axial (blue) and radial (orange) voltages respectively. The axial mode is not excited by the radial modulation that causes an even change in the vibrational quantum number. This measurement was performed for an endcap voltage of \SI{8}{\volt} and an estimated radio-frequency peak-to-peak voltage of \SI{380}{\volt}. }
    \label{fig:trap_modulation_scan}
\end{figure}

\begin{figure}[h]
    \centering
    \includegraphics[width=0.45 \textwidth]{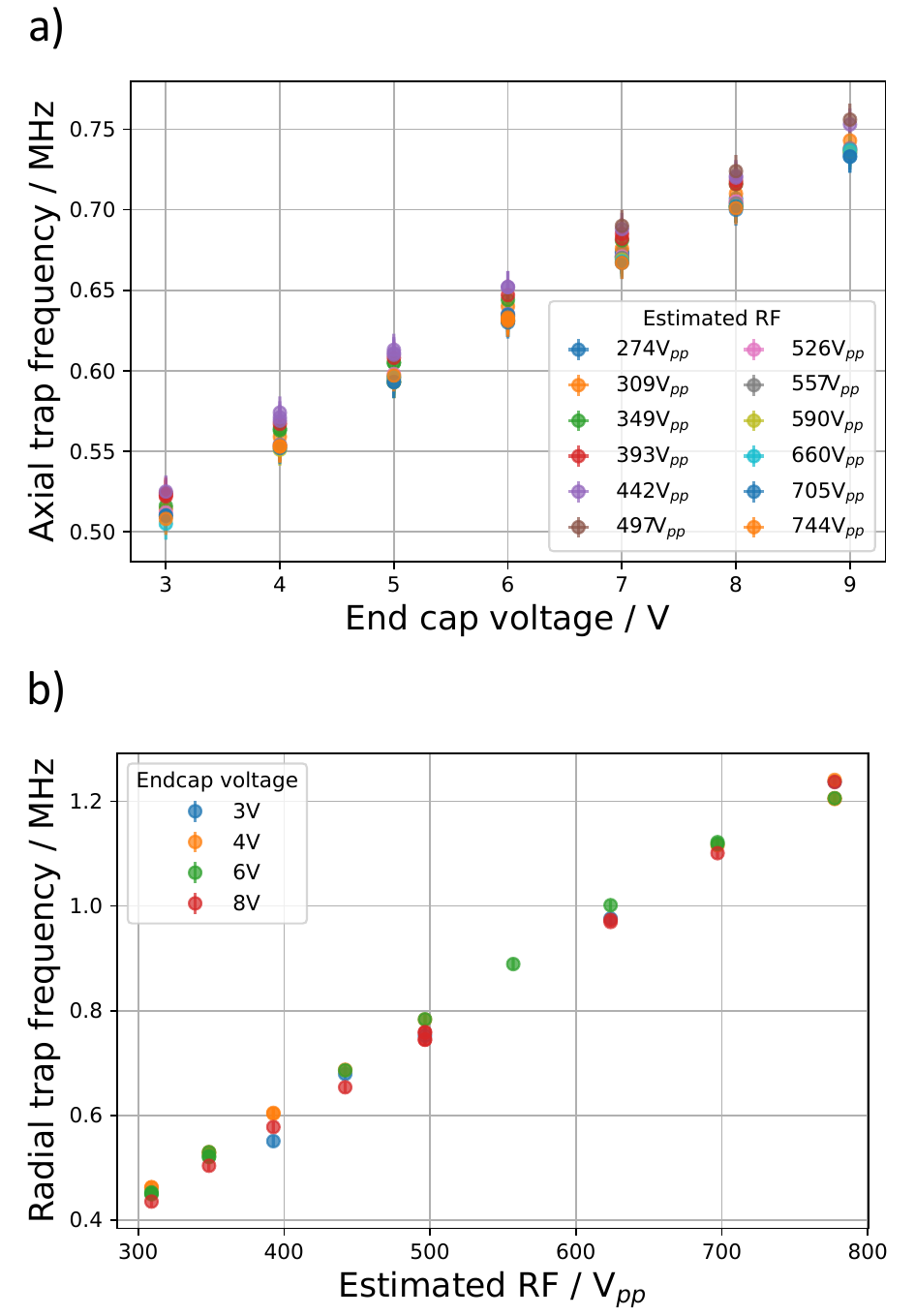}
    \caption {Trap frequency measurements as a function of the strength of the confinement. a) Axial trap frequency measurement. The voltage on one of the endcap electrodes was modulated to perform parametric excitation of axial modes. b) Radial trap frequency measurement. The radio-frequency drive of the ion trap was modulated to parametrically excite the radial modes. The peak-to-peak voltage on the linear electrodes was estimated from the two voltage pick-off antennas on the bias-tee board (see Figure \ref{fig:bias_tee_design}).}
    \label{fig:trap_frequency}
\end{figure}

% - Trap frequency measurements and how far we can reach
We measure the radial trap frequency by performing a parametric excitation using modulation of the radio-frequency voltage amplitude. With one ion at a frequency of $\omega_r \simeq 2\pi \times 1$\,MHz, the two radial modes are equal to within \SI{2}{\kilo\hertz}. In order to detect the axial modes, we add a small high-frequency signal to one of the end cap voltages. An example of axial and radial modulation scans is shown in figure \ref{fig:trap_modulation_scan}.  Modulation of the radio-frequency voltage amplitude excites the radial motional states with a change in the vibrational quantum number of $\Delta n = \text{even}$. Hence, the trap frequency corresponding is at half of the the value of the observed resonance frequency. In contrast, the axial modulation  causes a change of the vibrational quantum number of $\Delta n = \text{odd}$, implying the measured resonance frequency corresponds to the actual axial trap frequency.  Figure \ref{fig:trap_frequency} shows the results from trap-frequency measurements for different radio-frequency and end-cap voltages. Radial trap frequencies $\omega_r$  of up to $2\pi\cdot1.2$\si{\mega\hertz} have been observed. This corresponds to around $800$\,V$_\text{pp}$ on the radio-frequency electrodes from FEM simulation of the trap.

\section{Conclusion and outlook}

In conclusion, we have constructed and analysed a two-phase helical resonator for operating a radio-frequency ion trap. We successfully trap and cool a linear Coulumb crystal of Ytterbium ions in a short radio-frequency trap using gold coated optical fibers as endcaps. This is the first step towards building a compact fiber-coupled quantum-information node using a chain of trapped ions. As a next step, we plan to replace the fully coated fibers with a fiber cavity \cite{hunger_fiber_2010, steiner_single_2013} with a coated metal mask \cite{pfeifer_achievements_2022} allowing us to perform Cavity-QED experiments.

\begin{acknowledgments}

This work has been supported by the DFG (SFB/TR 185 project A2) and Cluster of Excellence Matter and Light for Quantum Computing (ML4Q) EXC 2004/1 – 390534769, project Quantenrepeater.Net (QR.N), and project FaResQ.

We thank F. Vewinger, W. Alt, J. Home, P. Schmitt, and F. Schmidt-Kaler for useful discussions regarding the work done here. We also thank S. Linden for the gold coating on the used optical fibers.

\end{acknowledgments}

\section*{Conflict of Interest}
The authors have no conflicts to disclose.

\section*{Data Availability Statement}

The raw data will available at 10.5281/zenodo.17950912.

\end{document}